\documentclass[pra,preprint,showpacs,superscriptaddress,amsfonts,amsmath,floatfix]{revtex4-1}

\usepackage{graphicx}
\usepackage{epstopdf}
\usepackage{bbold}
\usepackage{color}

\begin{document}


\title{Laser assisted tunneling in a Tonks-Girardeau gas}

\author{Karlo Lelas}
\affiliation{Faculty of Textile Technology, University of Zagreb, 
Prilaz baruna Filipovi\'{c}a 28a, 10000 Zagreb, Croatia}
\author{Nikola Drpi\'{c}}
\affiliation{Department of Physics, University of Zagreb, Bijeni\v{c}ka c. 32, 10000 Zagreb, Croatia}
\author{Tena Dub\v{c}ek}
\affiliation{Department of Physics, University of Zagreb, Bijeni\v{c}ka c. 32, 10000 Zagreb, Croatia}
\author{Dario Juki\'{c}}
\affiliation{Faculty of Civil Engineering, University of Zagreb, Fra Andrije Ka\v{c}i\'{c}a-Mio\v{s}i\'{c}a 26, 10000 Zagreb, Croatia}
\author{Robert Pezer}
\affiliation{Faculty of Metallurgy, University of Zagreb, Aleja narodnih heroja 3, HR 44103 Sisak, Croatia}
\author{Hrvoje Buljan}
\affiliation{Department of Physics, University of Zagreb, Bijeni\v{c}ka c. 32, 10000 Zagreb, Croatia}
\email{hbuljan@phy.hr}

\date{\today}

\begin{abstract}
We investigate the applicability of laser assisted tunneling in a strongly interacting 
one-dimensional Bose gas (the Tonks-Girardeau gas) in optical lattices. 
We find that the stroboscopic dynamics of the Tonks-Girardeau gas 
in a {\em continuous} Wannier-Stark-ladder potential, supplemented with laser assisted tunneling, 
effectively realizes the ground state of one-dimensional hard-core bosons in a 
{\em discrete} lattice with nontrivial hopping phases. 
We compare observables that are affected by the interactions, such as the 
momentum distribution, natural orbitals and their occupancies, 
in the time-dependent {\em continuous} system, to those of the 
ground state of the {\em discrete} system. 
Stroboscopically, we find an excellent agreement, indicating
that laser assisted tunneling is a viable technique for realizing 
novel ground states and phases with hard-core one-dimensional Bose gases.
\end{abstract}

\maketitle

\section{Introduction}

The fractional quantum Hall (FQH) state emerges in a system of strongly interacting charged particles in a
strong magnetic field and confined in two-dimensions~\cite{Tsui1982, Lau1983}.
The richness of this system motivates the quest for novel topological states of 
matter in other systems~\cite{Chen2013}. Ultracold atomic gases are an ideal playground 
for a controlled preparation, manipulation, and detection of quantum many-body 
states~\cite{Bloch2008}. However, to achieve topological states such as the FQH state 
in ultracold atomic gases, one must create a synthetic (artificial) magnetic field, 
wherein atoms behave as charged particles in magnetic fields~\cite{Dal2011,Bloch2012,Goldman2014}.

A variety of methods for the creation of synthetic magnetic fields have been 
implemented over the years~\cite{Madison2000,Abo2001,Lin2009,Aidelsburger2011,Struck2012,Miyake2013, Bloch2013}, 
including the Coriolis force method in rapidly rotating BECs~\cite{Madison2000,Abo2001}, 
and methods based on the Berry phase, which plays the role of the Aharon-Bohm phase (see ~\cite{Lin2009} for bulk BECs). 
In optical lattices, one engineers the amplitude and the phase of the tunneling matrix elements 
(hopping parameters)~\cite{Aidelsburger2011,Struck2012,Miyake2013, Bloch2013}, for example, by shaking the lattice \cite{Struck2012} or using laser assisted 
tunneling~\cite{Aidelsburger2011,Miyake2013,Bloch2013,Zoller2003,Kolovsky2011}. 
This has led to the experimental realization of paradigmatic condensed-matter 
Hamiltonians such as the Harper-Hofstadter Hamiltonian~\cite{Miyake2013,Bloch2013} and 
the Haldane Hamiltonian~\cite{Jotzu2014}. 
However, most of the efforts regarding synthetic magnetic fields were focused on 
single particle effects.

Recently, strong interactions were used in the physics of gauge fields, in the 
first observation of BEC (i.e. the ground state) in the Harper-Hofstadter 
Hamiltonian~\cite{Kennedy2015}. As synthetic magnetic fields in optical lattices 
are essentially obtained by periodic driving~\cite{Struck2012,Aidelsburger2011,Miyake2013}, 
an important question in this context is to understand the behavior of periodically 
driven interacting quantum systems~\cite{Alessio2013,Alessio2014,Bukov2015,Kuwahara2015,Eckardt2015}. 
From the eigenstate thermalization hypothesis it follows that 
driven interacting systems will, after sufficiently long time, heat up to an infinite 
temperature~\cite{Rigol2008}. However, in some regimes, the 
system can approach a prethermalized Floquet steady state before 
heating up~\cite{Bukov2015,Kuwahara2015}, implying that the method can be used
(in some regimes) with interactions present. 
Next, laser assisted tunneling was suggested as a scheme to 
engineer and promote three-body interactions in atomic gases~\cite{Daley2014}. 
An interplay of on-site Hubbard interactions and laser assisted tunneling 
was recently suggested for realizing versatile Hamiltonians in optical lattices~\cite{Bermudez2015}. 
Following the realization of the Harper-Hofstadter Hamiltonian with bosonic 
atoms~\cite{Miyake2013,Bloch2013}, two-dimensional strongly correlated lattice 
bosons in a strong magnetic field were recently studied~\cite{Natu2016}. 
In one-dimensional (1D) fermionic systems, the classification of topological 
phases was shown to depend on the presence/absence of interactions~\cite{FidkowskiPRB2010}.

In the quest for strongly correlated topological states, the applicability of methods 
for synthetic magnetic fields should be scrutinized in the presence of interactions. 
Here we examine the applicability of laser assisted tunneling~\cite{Aidelsburger2011,Miyake2013, Bloch2013} 
for a strongly interacting 1D Bose gas (the Tonks-Girardeau gas~\cite{Girardeau1960}) in 
an optical lattice. 
The Tonks-Girardeau model is exactly solvable via the Fermi-Bose mapping, i.e., by 
mapping a wave function for noninteracting spinless 1D fermions to that of 
impenetrable core 1D bosons~\cite{Girardeau1960}. The experimental realization of the Tonks-Girardeau 
gas in atomic waveguides, proposed by Olshanii~\cite{Olshanii}, has been acomplished 
more than a decade ago~\cite{gases1D1, gases1D2, gases1D3}. 
Impenetrable core interactions for bosons mimic the Pauli exclusion principle in $x$-space, 
thus, the single particle density is identical for the Tonks-Girardeau gas and 
noninteracting spinless fermions~\cite{Girardeau1960}. 
However, the two systems considerably differ in momentum space~\cite{Lenard1964}. 
The laser assisted tunneling should work well for noninteracting spinless fermions 
(on the Fermi side of the Fermi-Bose mapping~\cite{Girardeau1960}), but it is not 
immediately clear how will the interplay of this method and impenetrable core interactions 
affect the momentum distribution, and the other observables depending 
on phase coherence.

Here we demonstrate, by numerical calculations, that the stroboscopic dynamics of a 
Tonks-Girardeau gas in a {\em continuous} Wannier-Stark-ladder potential, 
supplemented with periodic driving, which simulates laser assisted tunneling, 
effectively realizes the ground 
state of hard-core bosons (HCB) on the {\em discrete} lattice with nontrivial hopping phases
(i.e. complex hopping parameters). 
We calculate the momentum distribution, natural orbitals and their occupancies for 
the ground state of HCB on such a discrete lattice, and find 
excellent agreement between these results, and observables 
calculated for a series of stroboscopic moments of the Tonks-Girardeau gas 
in continuous periodically driven Wannier-Stark-ladders. 

Before presenting our results, we further discuss the motivation for 
studying interacting 1D Bose gases in synthetic gauge fields. 
Quite generally, in addition to being exactly solvable (in some situations) and experimentally accessible, 
interacting 1D Bose gases present a many-body system with enhanced quantum effects 
due to the reduced dimensionality. 
More specifically, in discrete lattices with phase dependent hopping amplitudes, 
one may explore strongly correlated ground states and excitations with potentially 
intriguing many-body properties. It should be mentioned that 1D spin-polarized fermions, 1D hard-core bosons, 
and 1D hard-core anyons, are related through the Bose-Fermi and anyon-fermion 
mapping~\cite{Girardeau2006}. Free expansion of 1D hard core anyons has been studied 
in Ref.~\cite{delCampo2008}. In Ref.~\cite{delCampo2011}, 
the multi-particle tunneling decay (in 1D) was studied in dependence on  
interactions and statistics, by addressing it for these three types of 1D particles. 
Furthermore, laser-assisted tunneling addressed here, plays the key role in a recent  
proposal for the experimental realization of anyons in 1D optical lattices~\cite{Keilmann2011}.

\section{The Tonks-Girardeau model}

We consider a gas of N identical bosons in 1D, which interact via 
pointlike interactions, described by the Hamiltonian
\begin{equation}
\hat{H}=\sum_{i=1}^{N} \left[-\frac{\hbar^2}{2m}\frac{\partial^2}
{\partial x_{i}^2}+V(x_{i},t) \right]+g_{1D}\sum_{1\leq i<j \leq N}
\delta(x_i-x_j).
\label{Hamiltonian}
\end{equation}
Such a system can be realized with ultracold bosonic atoms trapped in 
effectively 1D atomic waveguides~\cite{gases1D1, gases1D2, gases1D3}, where 
$V(x)$ is the axial trapping potential, and 
\allowbreak{ 
$g_{1D}=2\hbar^2a_{3D}\left[ma^2_{\perp}(1-Ca_{3D}/\sqrt{2}a_{\perp})\right]^{-1}$}
is the effective 1D coupling strength; $a_{3D}$ stands for the three-dimensional 
$s$-wave scattering length, $a_{\perp}=\sqrt{\hbar/m\omega_{\perp}}$ is the transverse width of the 
trap, and $C=1.4603$~\cite{Olshanii}. By varying $\omega_{\perp}$, the system can be 
tuned from the mean field regime up to the strongly interacting Tonks-Girardeau 
regime~\cite{Girardeau1960} with infinitely repulsive contact interaction 
$g_{1D}\rightarrow\infty$. For the Tonks-Girardeau regime, the interaction term of the Hamiltonian 
(\ref{Hamiltonian}) can be replaced by a boundary condition on the many-body wave function~\cite{Girardeau1960}: 
\begin{equation}
\Psi_B(x_1,x_2,\ldots,x_N,t)=0\ \mbox{if}\ x_i=x_j
\label{conditionTG}
\end{equation}
for any $i\neq j$. With this boundary condition, the Hamiltonian becomes
\begin{equation}
\hat{H}=\sum_{i=1}^{N} \left[-\frac{\hbar^2}{2m}\frac{\partial^2}
{\partial x_{i}^2}+V(x_{i},t) \right].
\label{HamiltonianTG}
\end{equation}
The boundary condition (\ref{conditionTG}) and the Schr\"{o}dinger equation for 
(\ref{HamiltonianTG}) are satisfied by an antisymmetric many-body wave function 
$\Psi_F$ describing a system of noninteracting spinless fermions in 1D \cite{Girardeau1960}. 
Because the system is 1D, an exact (static and time-dependent) solution of the Tonks-Girardeau 
model can be written via the famous Fermi-Bose mapping~\cite{Girardeau1960}:
\begin{equation}
\Psi_B(x_1,x_2,\dots,x_N,t) = \prod_{1\leq i < j\leq N} \mbox{sgn}(x_i-x_j)\Psi_F(x_1,x_2,\dots,x_N,t)\;.
\label{mapFB}
\end{equation}
The fermionic wave function $\Psi_F$ can be written in the form of a Slater determinant 
(or generally as a superposition of such determinants),
\begin{equation}
\Psi_F(x_1,\ldots,x_N,t)=
\frac{1}{\sqrt{N!}} \det_{m,j=1}^{N} [\psi_m(x_j,t)], 
\label{psiF}
\end{equation} 
where $\psi_m(x,t)$ denotes $N$ orthonormal single particle  
wave functions obeying a set of uncoupled single-particle Schr\" odinger equations: 
\begin{equation}
i\hbar\frac{\partial \psi_m}{\partial t}=
\left [ - \frac{\hbar^2}{2m}\frac{\partial^2 }{\partial x^2}+
V(x,t) \right ] \psi_m(x,t), \ m=1,\ldots,N.
\label{master}
\end{equation}
Equations (\ref{mapFB})-(\ref{master}) prescribe the construction of
the many-body wave function describing the Tonks-Girardeau gas in an external 
potential $V(x,t)$. The mapping is applicable both in the stationary~\cite{Girardeau1960} and the
time-dependent case~\cite{Girardeau2000}. 

The expectation values of the one-body observables are obtained from the reduced 
single particle density matrix (RSPDM), defined as 
\begin{eqnarray}
\rho_{B}(x,y,t) & = &  N \int \!\! dx_2\ldots dx_N \, \Psi^*_B(x,x_2,\ldots,x_N,t)
\nonumber \\
&& \times \Psi_B(y,x_2,\ldots,x_N,t).\label{rspdm}
\end{eqnarray} 
Observables of interest here are the single particle $x$-density 
$\rho_{B}(x,x,t)=\sum_{m=1}^{N}|\psi_m(x,t)|^2$, and the momentum 
distribution \cite{Lenard1964}:
\begin{equation}
n_B(k,t) = \frac{1}{2\pi}\int \!\! dx dy \, e^{i k(x-y)}\rho_{B}(x,y,t). 
\label{MDformula}
\end{equation} 
A concept that is very useful for the understanding 
of the bosonic many-body systems is that of natural orbitals. 
The natural orbitals $\Phi_i(x,t)$ are eigenfunctions of 
the RSPDM, 

\begin{equation}
\int \!\! dx\, \rho_{B}(x,y,t) \, \Phi_i (x,t) =
\lambda_i(t) \, \Phi_i (y,t), \quad i=1,2,\ldots,
\label{NOs}
\end{equation}
where $\lambda_i$ are the corresponding eigenvalues;
the RSPDM is diagonal in the basis of natural orbitals,

\begin{equation}
\rho_{B}(x,y,t) = \sum_{i=1}^{\infty} 
\lambda_i(t) \Phi_i^* (x,t) \Phi_i (y,t).
\end{equation}
The natural orbitals can be interpreted as effective single-particle 
states occupied by the bosons, where $\lambda_i$ represents the 
occupancy of the corresponding orbital~\cite{Girardeau2001}. 
The fermionic RSPDM $\rho_F(x,y,t)$ and the momentum distribution 
$n_F(k)$ are defined by Eqs. (\ref{rspdm}) 
and (\ref{MDformula}) with $\Psi_B\rightarrow\Psi_F$.
The single particle density $\rho_{B}(x,x,t)$ is identical for the Tonks-Girardeau 
gas and the noninteracting Fermi gas~\cite{Girardeau1960}. 
However, the momentum distributions of the two systems on the two sides of the 
mapping considerably differ~\cite{Lenard1964}. 
The momentum distribution and $\rho_B(x,y,t)$ for the continuous Tonks-Girardeau 
model [Eqs. (\ref{conditionTG}) and (\ref{HamiltonianTG})] can be efficiently 
calculated by using the procedure outlined in Ref.~\cite{Pezer2007}.

\section{Laser assisted tunneling in a Tonks-Girardeau gas}

Our strategy is as follows: We study the ground state of hard core bosons (HCB) 
on a discrete lattice, with nontrivial phases of the hopping parameters 
(i.e. with complex hopping parameters). Then, we examine in detail the quantum dynamics of the 
Tonks-Girardeau gas in a continuous Wannier-Stark-ladder potential 
with periodic driving, which corresponds to laser assisted tunneling~\cite{Aidelsburger2011,Miyake2013, Bloch2013}. 
Parameters of the periodic drive are tuned to correspond to the 
phases of the hopping parameters in the discrete system. 
More specifically, we observe the (stroboscopic) quantum dynamics 
of the single particle density in $x$- and in $k$-space, and compare 
it with the ground state properties of HCB on the 
discrete lattice.

\begin{figure}[!h]
\centering
\includegraphics[scale=0.8]{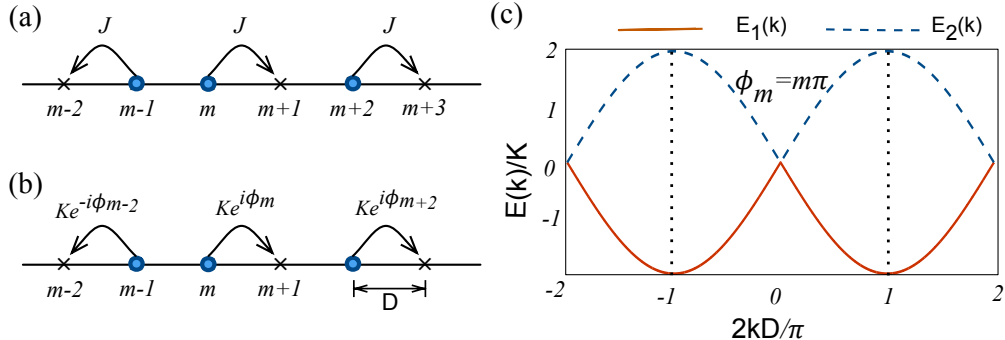} 
\caption{Discrete lattices and their energy bands.  
(a) The discrete lattice corresponding to the Hamiltonian (\ref{discrete1}) (hard core bosons are sketched as filled blue circles).   
(b) The discrete lattice with complex hopping parameters, which corresponds to 
the Hamiltonian (\ref{discrete2}). 
(c) Energy bands $E_{1,2}(k)$ of the discrete lattice illustrated in (b), with 
the phase of the hopping parameter $\phi_m=m\pi$. 
The edges of the first Brillouin zone are at $k=\pm\pi/2D$ 
(denoted with black vertical doted lines).
The energy bands are shown also outside of the first Brillouin zone 
for better visualization. 
} 
\label{Fig1}
\end{figure}

In our simulations of the Tonks-Girardeau gas with periodic driving, 
the gas is initially (at $t=0$) in the ground state of the optical 
lattice potential $V_L(x)$, which has $M=40$ lattice sites, and 
infinite wall boundary conditions:
\begin{align}
V_L(x)=
\begin{cases}
V_L \cos^2(\pi x/D), & \mbox{if } -20D\leq x\leq 20D \\
\infty, & \mbox{otherwise }.
\end{cases}
\label{lattice}
\end{align}
Here $V_L=10E_R$ is the amplitude of the optical lattice, $E_R=h^2/(2m\lambda^2)$ is the recoil 
energy, $\lambda=1.064\mu$m, and $D=\lambda/2$ is the period of the optical lattice. 
The lattice is loaded with $^{87}$Rb atoms, $m=1.455\times 10^{-25}$ kg. For such a deep lattice, 
the Tonks-Girardeau gas [described with Hamiltonian (\ref{HamiltonianTG}) together 
with condition (\ref{conditionTG})] can be approximated by the model of 
HCB on a discrete lattice~\cite{Rigol2004,Rigol2005}:  
\begin{equation}
\hat{H}=-J\sum_{m=1}^{M}\left[\hat{b}^{\dagger}_{m+1}\hat{b}_m+h.c\right],\quad 
\hat{b}^{\dagger 2}_{m}=\hat{b}^2_m=0,\quad
\left\lbrace \hat{b}_m,\hat{b}^{\dagger}_{m} \right\rbrace=1 
\label{discrete1}
\end{equation} 
Here, the bosonic creation and annihilation operators at site $m$ are 
denoted by $\hat{b}^{\dagger}_m$ and $\hat{b}_m$ respectively, 
and $J$ is the hopping parameter. The hard core constraint $\hat{b}^{\dagger 2}_m=\hat{b}^2_m=0$ precludes multiple occupancy of one lattice site, and the brackets in Eq. (\ref{discrete1}) apply only to on-site anticommutation relations, for $m\neq n$ these operators commute as usual for bosons $\left[\hat{b}_m,\hat{b}^{\dagger}_n\right]=0$. We calculate the effective hopping parameter $J=0.019 E_R$ from the Wannier states of the optical lattice by using the MLGWS code~\cite{Walters2013}. 
The discrete lattice of the Hamiltonian (\ref{discrete1}) is sketched in Fig.~\ref{Fig1}(a).     

In order to obtain the HCB discrete lattice model with tunable hopping amplitudes and phases, 
one can employ the laser assisted tunneling method. 
The scheme~\cite{Aidelsburger2011,Miyake2013, Bloch2013} utilizes far 
off-resonant lasers and a single atomic internal state, which minimizes 
heating by spontaneous emission. An early theoretical proposal related to this 
scheme was based on coupling of different internal states~\cite{Zoller2003}. 
The theoretical proposal in Ref.~\cite{Kolovsky2011} was later modified to
obtain a homogeneous synthetic magnetic field~\cite{Miyake2013, Bloch2013}. 
In order to simulate the scheme numerically, at $t=0$ we introduce the tilt potential 
$V_T(x)=\alpha x/D$, and simultaneously the time- and space-periodic potential 
$V_R(x,t)=V_R\cos^2\left[(q x-\omega t)/2\right]$. 
The periodic potential with the tilt, $V_L(x)+V_T(x)$, is the 
continuous Wannier-Stark-ladder potential; for a sufficiently large 
tilt, tunneling between neighboring lattice sites is suppressed~\cite{Miyake2013, Bloch2013}. 
The periodic drive potential $V_R(x,t)$ simulates two-photon Raman transitions
used to restore the tunneling~\cite{Miyake2013, Bloch2013}, and introduce  
nontrivial phases in the hopping parameters. 
Thus, for $t \geq 0$, the Tonks-Girardeau gas evolves in the time-periodic potential 
\begin{align}
V(x,t)=
\begin{cases}
V_L (x)+V_T(x)+V_R(x,t), & \mbox{if } -20D\leq x\leq 20D \\
\infty, & \mbox{otherwise }.
\end{cases}
\label{totalpot}
\end{align}
The strength of the tilt and the drive are set by $\alpha=0.1V_L$ and $V_R=0.17V_L$, respectively. 
The frequency of driving is in resonance with the energy offset between neighboring sites 
of the tilted potential, that is, $\omega=\alpha/\hbar$.

For a deep optical lattice, our continuous model with potential (\ref{totalpot}) can be approximated 
by a discrete Hamiltonian with the kinetic (hopping) term, tilt, drive, and on-site interactions.
Such a discrete Hamiltonian is a starting point in Refs.~\cite{Polkovnikov2014, GoldmanPRA2015}, 
for deriving a discrete model with complex hopping amplitudes and interactions.
More specifically, in the case of resonant driving ($\omega=\alpha/\hbar$), a unitary 
transformation can cast this discrete Hamiltonian into a rotating frame, such that the 
kinetic term, together with the tilt and drive terms, become an effective  
kinetic term with complex hopping amplitudes~\cite{Polkovnikov2014, GoldmanPRA2015}. 
The on-site interaction term is not affected by this unitary transformation~\cite{Polkovnikov2014, GoldmanPRA2015}. 
The derivation is applicable for any strength of the interaction $U/J$~\cite{Polkovnikov2014, GoldmanPRA2015}, where $U$ stands 
for the on-site interaction energy. Even though the derivation 
in Refs. \cite{Polkovnikov2014, GoldmanPRA2015} is for 2D lattices, it is applicable for 1D lattices as well; 
it is also valid in the strongly interacting limit $U/J\rightarrow\infty$ studied here. 
In other words, the Tonks-Girardeau gas in the continuous potential (\ref{totalpot}) can be approximated 
with the Hamiltonian of hard core bosons (HCB) on a discrete lattice with nontrivial hopping phases: 
\begin{equation}
\hat{H}=-K\sum_{m=1}^{M}\left[e^{i\phi_m}\hat{b}^{\dagger}_{m+1}\hat{b}_m+h.c\right],
\quad \hat{b}^{\dagger 2}_{m}=\hat{b}^2_m=0,\quad
\left\lbrace \hat{b}_m,\hat{b}^{\dagger}_{m} \right\rbrace=1,
\label{discrete2}
\end{equation} 
where the hopping phase is given with $\phi_m=q D m$ and $K$ is the effective hopping amplitude. By choosing a different gauge, the phases in Eq. (\ref{discrete2}) can be eliminated. Nevertheless, a comparison of the continuous Tonks-Girardeau system in the time-dependent potential (\ref{totalpot}), with the discrete model (14), provides valuable information on the applicability of laser assisted tunneling in the presence of strong interactions. Moreover, these results have implications for interacting systems where nontrivial phases cannot be eliminated by a gauge transformation, and in systems with gauge dependent observables.  
We consider a discrete lattice with $M=40$ sites, corresponding to $M=40$ lattice sites 
of the optical lattice potential (\ref{lattice}). 
The discrete lattice of the Hamiltonian (\ref{discrete2}) is sketched in Fig.~\ref{Fig1}(b).

In what follows, we make a particular choice of the phases of the hopping parameters. 
In experiments, this is set by choosing the angle between the Raman beams~\cite{Miyake2013, Bloch2013}, 
and here we set it by choosing $q=\pi/D$ in the time-dependent potential 
$V_R(x,t)$; thus the hopping phase is $\phi_m=m\pi$. 
For this choice of the hopping phase, the discrete lattice has alternating hopping 
matrix elements, $K(-1)^m$, for tunneling from site $m$ to site $m+1$. 
We estimate the effective hopping amplitude to be $K=0.012E_R$. 
This is obtained by comparing the expansion of the initially localized 
single particle Gaussian wave packet in the total potential (\ref{totalpot}), with 
the expansion in the discrete lattice (\ref{discrete2}), and adjusting $K$ until the 
two patterns coincide; this method was adopted from Ref. \cite{Tena2015}.

In order to obtain the ground state properties of the HCB Hamiltonian (\ref{discrete2}), 
we use the Jordan-Wigner transformation~\cite{JWT, Rigol2004, Rigol2005}
\begin{equation}
\hat{b}^{\dagger}_m=\hat{f}^{\dagger}_m\prod^{m-1}_{\beta=1}e^{-i\pi \hat{f}^{\dagger}_{\beta}\hat{f}_{\beta}}, \quad \hat{b}_m=\prod^{m-1}_{\beta=1}e^{i\pi \hat{f}^{\dagger}_{\beta}\hat{f}_{\beta}}\hat{f}_m ,
\label{JW}
\end{equation}
which maps the HCB Hamiltonian (\ref{discrete2}), to the Hamiltonian for 
discrete noninteracting spinless fermions:
\begin{equation}
\hat{H}=-\sum_{m=1}^{M}\left[Ke^{i\phi_m}\hat{f}^{\dagger}_{m+1}\hat{f}_m+h.c\right],
\label{fermions}
\end{equation}   
where $\hat{f}^{\dagger}_m$ and $\hat{f}_m$ are the creation and annihilation operators for 
spinless fermions. We calculate the ground state momentum distribution $N_B(k)$ of 
the HCB Hamiltonian (\ref{discrete2}), and the ground state momentum distribution 
$N_F(k)$ of the noninteracting spinless fermions Hamiltonian (\ref{fermions}), 
by using the procedure outlined by Rigol and Muramatsu~\cite{Rigol2004, Rigol2005}.    

\begin{figure}[!h]
\centering
\includegraphics[scale=0.45]{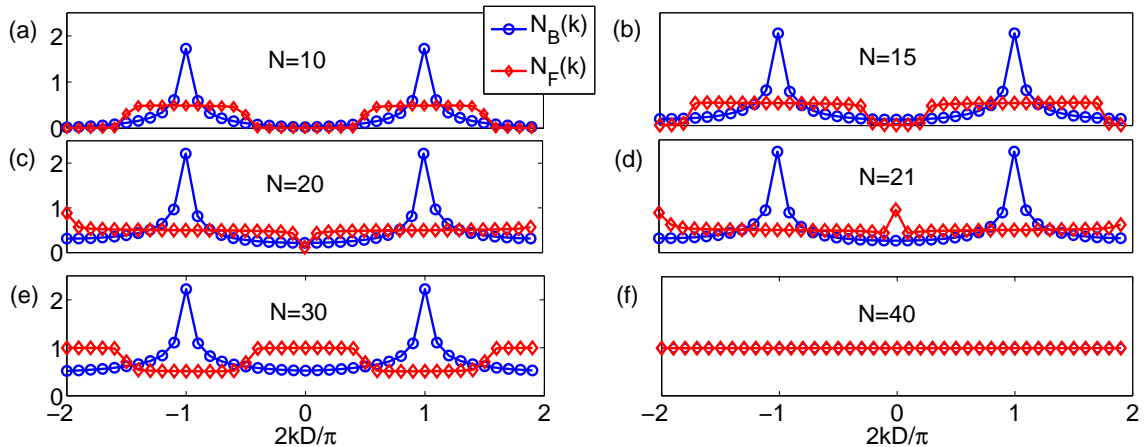} 
\caption{Ground state momentum distribution $N_B(k)$ (blue circles) of hard core bosons and 
noninteracting spinless fermions $N_F(k)$ (red diamonds) in the discrete lattice with the 
hopping phase $\phi_m=m\pi$ [see Figs.~\ref{Fig1}(b) and (c)]. 
Illustrated are cases for the number of particles 
$N=10,15,20,21,30$, and $40$.  For $N=20$ the first band $E_1(k)$ is filled, and 
the second band $E_2(k)$ is empty. For $N=40$ the bands are filled, and the system is 
in the Mott state of one atom per lattice site with $N_B(k)=N_F(k)$. 
The bosonic distribution $N_B(k)$ is for all values of $N$ peaked at 
$k=\pm\pi/2D$; these values are at the edges of the first Brillouin zone, that is, 
at the minima of the first energy band $E_1(k)$.
In contrast, the fermionic distribution has a characteristic Fermi plateau, which 
becomes inverted when the second band starts to be filled for $N>M/2=20$. 
See text for details.}
\label{Fig2}
\end{figure}

In Figs.~\ref{Fig2}(a)-(f) we show $N_B(k)$ and $N_F(k)$ in dependence on 
the number of particles $N$. 
The figure can be understood by considering the single-particle energy bands 
of the discrete lattice with $\phi_m=m\pi$, illustrated in Fig.~\ref{Fig1}(c). 
There are two bands, $E_1(k)=-2K|\sin(kD)|$ and $E_2(k)=2K|\sin(kD)|$, which touch at a 
1D Dirac point at $k=0$. 
The ground state of $N$ hard core bosons is constructed by using the first $N$ 
single particle states~\cite{Rigol2004, Rigol2005}, which on the Fermi side of the 
mapping fill the states up to the Fermi level. Note that for $N=20$, the band 
$E_1(k)$ is filled and $E_2(k)$ empty, while for $N=40$ they are both full. 
For $N<M/2=20$, the single particle states  
partially fill the first band $E_1(k)$ which has minima at the edges of the first 
Brillouine zone at $\pm\pi/2D$; thus both $N_B(k)$ and $N_F(k)$ are centered at these 
values. The fermionic distribution has the characteristic plateau(s) with Fermi edges, 
whereas the bosonic distribution has a spike at the maxima; the spike is a consequence of the 
fact that bosons tend to occupy the same single particle state of lowest energy. 
For filling $M/2<N<M=40$, the bosonic momentum distribution 
retains its peak at the edges of the Brillouin zone $\pm\pi/2D$, whereas the 
fermionic plateau is inverted because the Fermi level is now at the 2nd band $E_2(k)$. 
In Fig.~\ref{Fig2}(f) we show the results for filled bands $N=M=40$, i.e. both 
bosons and fermions are in the Mott insulating state with one
particle per lattice site and the two momentum distributions overlap, $N_B(k)=N_F(k)$.    

\begin{figure}[!h]
\centering
\includegraphics[scale=0.4]{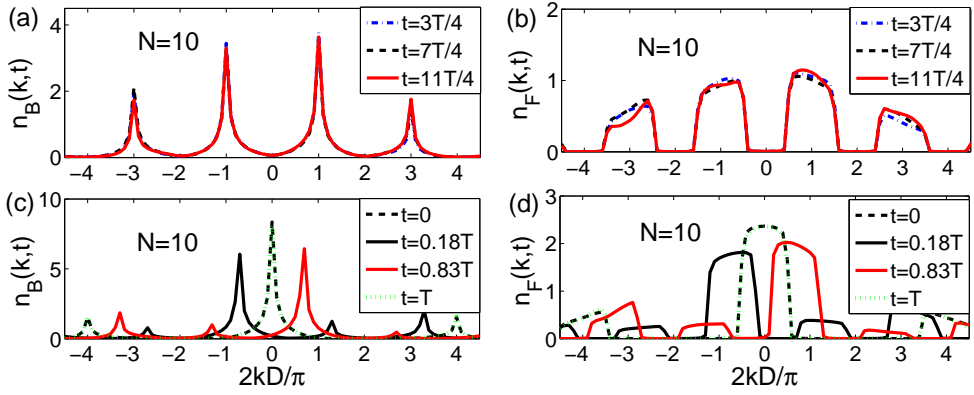}
\caption{Time dependence of the momentum distribution $n_B(k,t)$ for 
Tonks-Girardeau bosons in the continuous Wannier-Stark-ladder potential
with periodic driving, and $n_F(k,t)$ for free fermions, for $N=10$ particles. 
(a) $n_B(k,t)$ at stroboscopic instants $t=\lbrace3T/4,7T/4,11T/4\rbrace$ (the lines overlap). 
Results are in agreement with the momentum distribution $N_B(k)$ for 
the discrete HCB model [blue circles in Fig.~\ref{Fig2}(a)].  
(b) $n_F(k,t)$ at stroboscopic instants $t=\lbrace3T/4,7T/4,11T/4\rbrace$ (the lines overlap).
(c)-(d) $n_B(k,t)$ and $n_F(k,t)$ at non-stroboscopic instants $t=\lbrace0,0.18T,0.83T,T\rbrace$. 
See text for details.}
\label{Fig3}
\end{figure}

Next, we explore the quantum dynamics of the Tonks-Girardeau gas 
in the {\em continuous} Wannier-Stark-ladder potential $V_L(x)+V_T(x)$
with periodic driving $V_R(x,t)$ (simulating laser assisted tunneling). 
The initial state is the ground state of the optical lattice potential 
$V_L(x)$ (\ref{lattice}). Note that we use capital letters ($N_B(k)$) to describe momentum distributions 
of the discrete model, and lower case ($n_B(k)$) for the momentum distribution of the continuous model.
We find that the time dependent momentum distribution $n_B(k,t)$ 
for Tonks-Girardeau bosons, and $n_F(k,t)$ for free fermions, are {\em stroboscopically} \cite{Polkovnikov2014}
in excellent agreement with the momentum distributions $N_B(k)$ and $N_F(k)$ of the discrete lattice 
models [Hamiltonians (\ref{discrete2}) and (\ref{fermions})]. 
More specifically, at times $t=(4n-1)T/4$, $n=\lbrace1,2,3...\rbrace$, 
where $T=2\pi/\omega=0.497$ ms is the period of the periodic driving, 
the {\em continuous} and {\em discrete} model momentum distributions coincide. 
The $N_B(k)$ of the discrete model has its domain in the 
first Brillouin zone $k\in\left[-\pi/2D,\pi/2D\right]$, whereas 
for $n_B(k,t)$ of the continuous model $k\in\left(-\infty,\infty\right)$. 
Apart from the main peaks in the 1$^{st}$ Brillouin zone, 
the continuous momentum distributions $n_B(k,t)$ have additional (expected) 
peaks at positions shifted by an integer number of reciprocal lattice vectors.

In Figs.~\ref{Fig3}(a) and (b) we show $n_B(k,t)$ and $n_F(k,t)$ for $N=10$ particles 
at the first three stroboscopic times $t=\lbrace3T/4,7T/4,11T/4\rbrace$; the lines overlap
indicating that a Floquet steady state is reached. 
We see excellent agreement of the continuous momentum distributions $n_B(k,t)$ and $n_F(k,t)$, 
presented in Figs.~\ref{Fig3}(a) and (b), with the momentum distributions 
$N_B(k)$ and $N_F(k)$ of the discrete model for $N=10$ particles, Fig.~\ref{Fig2}(a).

In Figs.~\ref{Fig3}(c) and (d) we show the momentum distributions $n_B(k,t)$ and $n_F(k,t)$ 
for non-stroboscopic moments for $N=10$ particles. At $t=0$, the Tonks-Girardeau 
gas is in the ground state of the optical lattice (\ref{lattice}), which can be 
approximated with the discrete model (\ref{discrete1}) sketched in Fig.~\ref{Fig1}(a). 
The discrete model (\ref{discrete1}) has one energy band $E(k)=-2J\cos(kD)$. 
In Fig.~\ref{Fig3}(c) we show the momentum distributions $n_B(k,t)$ of the initial state at 
$t=0$ (black doted line) and at $t=T$ (green doted line); we see that the maxima of 
$n_B(k,t)$ are at $k=0,\pm4\pi/2D$, which is consistent with the minima of the band $E(k)$. 
The same reasoning holds for $n_F(k,t)$ at times $t=\lbrace0,T\rbrace$ in Fig.~\ref{Fig3}(d). 
For completeness, in Fig.~\ref{Fig3}(c), we also show $n_B(k,t)$ for non-stroboscopic times $t=\lbrace0.18T,0.83T\rbrace$ 
(black and red solid lines respectively); at these times momentum distributions 
$n_B(k,t)$ have one dominant maximum and several smaller 
peaks at various $k$.  
       
\begin{figure}[!h]
\centering
\includegraphics[scale=0.4]{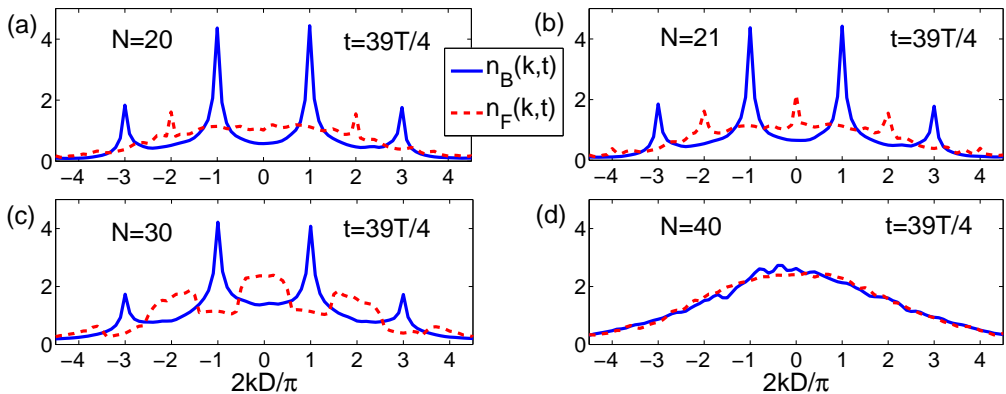}
\caption{Momentum distributions $n_B(k,t)$ (blue solid line) for 
Tonks-Girardeau bosons in the continuous Wannier-Stark-ladder potential
with periodic driving, and $n_F(k,t)$ (red dashed line) for free fermions, 
in dependence of the number of particles $N$, at the 10th stroboscopic 
instant $t=39T/4\approx5$ ms. 
The results are in agreement with momentum distributions $N_B(k)$ and 
$N_F(k)$ of the discrete models presented in Figs.~\ref{Fig2}(c)-(f). 
See text for details.}  
\label{Fig4}
\end{figure}

In Figs.~\ref{Fig4}(a)-(d), we show the momentum distributions $n_B(k,t)$ and $n_F(k,t)$ at the 
10th stroboscopic appearance $t=39T/4\approx5$ ms for different numbers of particles, $N=\lbrace20,21,30,40\rbrace$.
We find excellent agreement with the momentum distributions $N_B(k)$ and $N_F(k)$ 
[shown in Figs.~\ref{Fig2}(c)-(e)] of the discrete Hamiltonians (\ref{discrete2}) 
and (\ref{fermions}), respectively. For $N=40$, the momentum distributions $n_B(k,t)$ and $n_F(k,t)$ do not have sharp peaks, and they slowly decay as $k\rightarrow\infty$; this shape differs from the uniform $N_B(k)$ and $N_F(k)$ corresponding to the Mott state shown in Fig.~\ref{Fig2}(f), which is expected for the lattice of finite depth. However, the bosonic and fermionic distributions coincide both in the continuous and the discrete model.   

\begin{figure}[!h]
\centering
\includegraphics[scale=0.4]{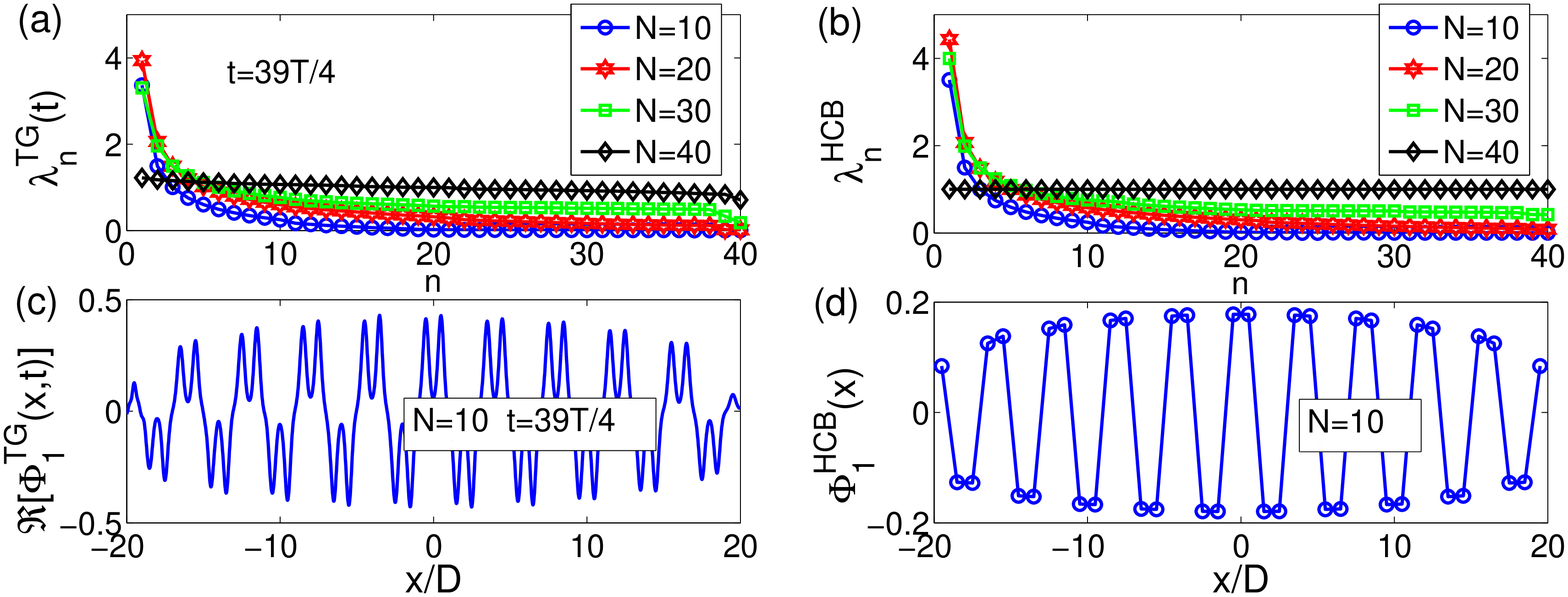}
\caption{Natural orbitals and their occupancies. (a) The occupancies of natural 
orbitals of Tonks-Girardeau bosons in the continuous Wannier-Stark-ladder potential
with periodic driving at the 10th stoboscopic appearence $t=39T/4\approx5$ ms, for 
$N=\lbrace10,20,30,40\rbrace$ particles. 
(b) The occupancies of natural orbitals for the ground state of the hard core bosons 
(HCB) on a discrete lattice with the hopping phase $\phi_m=m\pi$. 
(c) Spatial dependence of the real part of the first natural orbital for the Tonks-Girardeau gas (continuous model).
(d) The first natural orbital of HCB bosons (discrete model). 
Both orbitals in (c) and (d) have the signature of the phase of the hopping parameter. 
See text for details.}
\label{Fig5}
\end{figure}

Besides momentum distributions, it is instructive to study the dynamics of natural orbitals 
$\Phi^{TG}_n(x)$ defined in Eq. (\ref{NOs}), and their occupancies $\lambda^{TG}_n$. 
In Fig.~\ref{Fig5}(a), we show $\lambda^{TG}_n$ for the continuous model at 
$t=39T/4$, for $N=\lbrace10,20,30,40\rbrace$ particles respectively; 
as expected, bosons partially condense in the first natural orbital 
$\Phi^{TG}_1(x)$, and this is reflected in the fact that the peak 
in their momentum distribution is located at the minima of the 
bands irrespective of the filling, which is not the case for fermions [see Fig. \ref{Fig1}]. 
For $N=40$, bosons are in the Mott state with $\lambda^{TG}_n\approx 1$ for every $n$. 
In Fig.~\ref{Fig5}(b), we show occupancies $\lambda^{HCB}_n$ of the natural orbitals 
$\Phi^{HCB}_n(x)$ within the discrete model (calculated with the procedure outlined 
in Ref.~\cite{Rigol2004,Rigol2005}). 
We find the same behavior as for the occupancies $\lambda^{TG}_n$ shown in 
Fig.~\ref{Fig5}(a). 

In Fig.~\ref{Fig5}(c), we show the real part of the first natural orbital $\Phi^{TG}_1(x,t)$ for 
$N=10$ at $t=39T/4$; we see the signature of the effective hopping phase 
$\phi_m=m\pi$, i.e. the real part of the first orbital changes the sign every two sites (the imaginary part follows the same behavior), and 
higher natural orbitals follow this behavior. For comparison, in 
Fig.~\ref{Fig5}(d) we show the first natural orbital $\Phi^{HCB}_1(x)$ of the ground state for 
$N=10$ HCB on the discrete lattice. We see excellent agreement between the two models. 

\section{Conclusion and outlook}

In conclusion, we investigated the applicability of the laser assisted tunneling in 
a strongly interacting Tonks-Girardeau gas. We found that the stroboscopic dynamics 
of the Tonks-Girardeau gas with laser assisted tunneling effectively realizes 
the ground state of 1D hard core bosons on a discrete lattice with nontrivial 
hopping phases. 
Our strategy was to compare the quantum dynamics of the Tonks-Girardeau gas 
in a continuous Wannier-Stark-ladder potential with 
periodic driving, which simulated laser assisted tunneling, 
with the ground state of the discrete model. 
More specifically, we have compared the momentum distribution, natural orbitals and their 
occupancies at stroboscopic moments corresponding to the period of the driving potential, 
and found excellent agreement. 

In the outlook, we would like to point out that by comparing continuous and discrete 
systems (see also \cite{Tena2015}), as we have done here, one opens the way for the study of shallow optical lattices 
with periodic driving, where the tight-binding approximation is not applicable. 
These continuous systems could potentially be described as discrete models with 
the phase dependent next to nearest neighbors hopping amplitudes (multi-band models). 
It is reasonable to expect that such models, with interactions present, possess many 
intriguing quantum states yet to be explored. Next, we would like to point out that a similar study could be in principle performed for interactions of finite strength, where one could explore modulations of the interaction strength \cite{Santos2012,Nagerl2016} leading to systems with density dependent hopping parameters.

\acknowledgments

This work was supported by the QuantiXLie Center of Excellence.


\begin{thebibliography}{99}

\bibitem{Tsui1982}
D. C. Tsui, H. L. Stormer, and A. C. Gossard,
Phys. Rev. Lett. {\bf 48}, 1559 (1982).

\bibitem{Lau1983}
R. B. Laughlin,
Phys. Rev. Lett. {\bf 50}, 1395 (1983).

\bibitem{Chen2013}
X. Chen, Z.-C. Gu, Z.-X. Liu, X.-G. Wen, 
arXiv:1301.0861 [cond-mat.str-el]; Science
338, 1604 (2012).


\bibitem{Bloch2008}
I. Bloch, J. Dalibard, and W. Zwerger,
Rev. Mod. Phys. {\bf 80}, 885 (2008).

\bibitem{Dal2011}
J. Dalibard, F. Gerbier, G. Juzeliunas, P. \"{O}hberg, 
Rev. Mod. Phys. {\bf 83}, 1523 (2011).

\bibitem{Bloch2012}
I. Bloch, J. Dalibard, and S. Nascimbene,
Nature Physics {\bf 8}, 267 (2012). 

\bibitem{Goldman2014}
N. Goldman, G. Juzeliunas, P. Ohberg, I. B. Spielman,
Rep. Prog. Phys. 77, 126401 (2014).


\bibitem{Madison2000}
K.W. Madison, F. Chevy, W. Wohlleben, and J. Dalibard,
Phys. Rev. Lett. {\bf 84}, 806 (2000). 

\bibitem{Abo2001}
J.R. Abo-Shaeer, C. Raman, J.M. Vogels, and W. Ketterle, 
Science {\bf 292}, 476 (2001).

\bibitem{Lin2009}
Y-J. Lin, R.L. Compton, K. Jim\'{e}nez-Garc\'{i}a, J.V. Porto, I.B. Spielman, 
Nature {\bf 462}, 628 (2009).

\bibitem{Aidelsburger2011}
M. Aidelsburger, M. Atala, S. Nascimbene, S. Trotzky, Y.-A. Chen, and I. Bloch, 
Phys. Rev. Lett. {\bf 107}, 255301 (2011).

\bibitem{Struck2012}
J. Struck, C. \"{O}lschl\"{a}ger, M. Weinberg, P. Hauke, J. Simonet, 
A. Eckardt, M. Lewenstein, K. Sengstock, and P. Windpassinger,
Phys. Rev. Lett. {\bf 108} 225304 (2012). 

\bibitem{Miyake2013}
H. Miyake, G.A. Siviloglou, C.J. Kennedy, W. Cody Burton, and W. Ketterle,
Phys. Rev. Lett. {\bf 111}, 185302 (2013).

\bibitem{Bloch2013}
M. Aidelsburger, M. Atala, M. Lohse, J. T. Barreiro, B. Paredes, and I. Bloch,
Phys. Rev. Lett. {\bf 111}, 185301 (2013).

\bibitem{Zoller2003}
D. Jaksch and P. Zoller
New J. Phys. {\bf 5} 56 (2003).

\bibitem{Kolovsky2011}
A. R. Kolovsky,
Europhys. Lett. {\bf 93} 003 (2011).

\bibitem{Jotzu2014}
G. Jotzu, M. Messer, R\'{e}mi Desbuquois, M. Lebrat, T. Uehlinger,
D. Greif, and T. Esslinger, Nature {\bf 515}, 237 (2014). 

\bibitem{Kennedy2015}
C.J. Kennedy, W. Cody Burton, Woo Chang Chung, and W. Ketterle
Nature Physics {\bf 11}, 859 (2015). 



\bibitem{Alessio2013}
L. D'Alessio and A.Polkovnikov, 
Ann. Phys. {\bf 333}, 19 (2013).

\bibitem{Alessio2014}
L. D'Alessio and M. Rigol, Phys. Rev. X {\bf 4}, 041048 (2014).

\bibitem{Bukov2015}
M. Bukov, S. Gopalakrishnan, M. Knap, and E. Demler,
Phys. Rev. Let. {\bf 115}, 205301 (2015) 

\bibitem{Kuwahara2015}
T. Kuwahara, T. Mori, and K. Saito, 
Ann. Phys. {\bf 367}, 96 (2015)

\bibitem{Eckardt2015}
A. Eckardt and E. Anisimovas, New J. Phys. {\bf 17}, 093039 (2015). 

\bibitem{Rigol2008}
M. Rigol, V. Dunjko, and M. Olshanii, Nature (London)
{\bf 452}, 854 (2008).

\bibitem{Daley2014}
A.J. Daley and J. Simon, Phys. Rev. A {\bf 89} 053619 (2014).

\bibitem{Bermudez2015}
A. Bermudez and D. Porras, 
New J. Phys. {\bf 17}, 103021 (2015).

\bibitem{Natu2016}
S.S. Natu, E.J. Mueller, and S. Das Sarma,
Phys. Rev. A {\bf 93}, 063610 (2016)

\bibitem{FidkowskiPRB2010}
L. Fidkowski and A. Kitaev,
Phys. Rev. B {\bf 81}, 134509 (2010). 


\bibitem{Girardeau1960}
M. Girardeau,
J. Math. Phys. {\bf 1}, 516 (1960). 

\bibitem{Olshanii}
M. Olshanii,
Phys.Rev.Lett. {\bf 81}, 938 (1998).

\bibitem{gases1D1}
T. Kinoshita, T. Wenger, and D. S. Weiss, 
Science {\bf 305}, 1125 (2004);

\bibitem{gases1D2}
B. Paredes, A. Widera, V. Murg, O. Mandel, S. F\"{o}lling,
I. Cirac, G. V. Shlyapnikov, T. W. H\"{a}nsch, and I. Bloch,
Nature (London) {\bf 429}, 277 (2004).

\bibitem{gases1D3}
T. Kinoshita, T. Wenger, and D. S. Weiss,
Nature (London) {\bf 440}, 900 (2006).

\bibitem{Lenard1964}
A. Lenard, J. Math. Phys. {\bf 5}, 930 (1964);
T.D. Schultz, J. Math. Phys. {\bf 4}, 666 (1963);
H.G. Vaidya and C.A. Tracy,
Phys. Rev. Lett. {\bf 42}, 3 (1979).

\bibitem{Girardeau2006}
M.D. Girardeau,
Phys. Rev. Lett. {\bf 97}, 100402 (2006).

\bibitem{delCampo2008}
A. del Campo, 
Phys. Rev. A {\bf 78}, 045602 (2008). 

\bibitem{delCampo2011}
A. del Campo,
Phys. Rev. A {\bf 84}, 012113 (2011). 

\bibitem{Keilmann2011}
T. Keilmann, S. Lanzmich, I. McCulloch, and M. Roncaglia, 
Nature Communications {\bf 2}, 361 (2011).

\bibitem{Girardeau2000}
M.D. Girardeau and E.M. Wright,
Phys. Rev. Lett. {\bf 84}, 5691 (2000). 

\bibitem{Girardeau2001}
M.D. Girardeau, E.M. Wright, and J.M. Triscari, 
Phys. Rev. A {\bf 63}, 033601 (2001); 
G.J. Lapeyre, M.D. Girardeau, and E.M. Wright, 
Phys. Rev. A {\bf 66}, 023606 (2002).

\bibitem{Pezer2007}
R. Pezer and H. Buljan, 
Phys. Rev. Lett. {\bf 98}, 240403 (2007).

\bibitem{JWT}
P. Jordan and E. Wigner, 
Z. Phys. {\bf 47}, 631 (1928).

\bibitem{Rigol2004}
M. Rigol and A. Muramatsu, 
Phys. Rev. Lett. {\bf 93}, 230404 (2004).

\bibitem{Rigol2005}
M. Rigol and A. Muramatsu, 
Mod. Phys. Lett. B {\bf 19}, 861 (2005).

\bibitem{Walters2013}
R. Walters, G. Cotugno, T. H. Johnson, S. R. Clark, and D. Jaksch,
Phys. Rev. A {\bf 87}, 043613 (2013).

\bibitem{Polkovnikov2014}
M. Bukov and A. Polkovnikov,
Phys. Rev. A {\bf 90}, 043613 (2014).

\bibitem{GoldmanPRA2015}
N. Goldman, J. Dalibard, M. Aidelsburger, and N.R. Cooper,
Phys. Rev. A {\bf 91}, 033632 (2015).

\bibitem{Tena2015}
T. Dub\v{c}ek, K. Lelas, D. Juki\'{c}, R. Pezer, M. Solja\v{c}i\'{c}, and H. Buljan,
New J. Phys. {\bf 17}, 125002 (2015).

\bibitem{Santos2012}
\'{A}. Rapp, X. Deng, and L. Santos, 
Phys. Rev. Lett. {\bf 109}, 203005 (2012).

\bibitem{Nagerl2016}
F. Meinert, M.J. Mark, K. Lauber, A.J. Daley, and H.C. Nägerl,
Phys. Rev. Lett. {\bf 116}, 205301 (2016).

\end{thebibliography}
\end{document}